\documentclass[prd,nofootinbib,showpacs,preprint,superscriptaddress,floatfix]{revtex4}

\usepackage{amsmath,amssymb}
\usepackage{graphicx}

\newcommand{\eVq}  {\text{eV}^2}
\newcommand{\Atm}  {\textsc{atm+reac}}
\newcommand{\Fut}  {\textsc{atm+reac+lbl}}
\newcommand{\Chooz}{\textsc{chooz}}

\newcommand{\Dmq}    {\Delta m^2}
\newcommand{\BDmq}   {\Delta\overline{m}^2}

\newcommand{\Btheta} {\overline{\theta}}

\newcommand{\Bomega} {\overline{\omega}}

\newcommand{\SKtim}[1] {SK$\times${#1}}

\allowdisplaybreaks

\begin{document}


\title{Measuring the deviation of the 2-3 lepton mixing from maximal
   with atmospheric neutrinos}

\preprint{hep-ph/0408170}
\preprint{YITP-SB-04-42}
\preprint{CERN-PH-TH/2004-156}

\author{M.~C.~Gonzalez-Garcia}
\email{concha@insti.physics.sunysb.edu}
\affiliation{Physics Department, 
  Theory Division, CERN, CH-1211, Geneva 23, Switzerland}
\affiliation{C.N.~Yang Institute for Theoretical Physics,
  SUNY at Stony Brook, Stony Brook, NY 11794-3840, USA}
\affiliation{IFIC, Universitat de Val\`encia -- C.S.I.C., Apt.~22085,
  E-46071 Val\`encia, Spain}

\author{M.~Maltoni}
\email{maltoni@insti.physics.sunysb.edu}
\affiliation{C.N.~Yang Institute for Theoretical Physics,
  SUNY at Stony Brook, Stony Brook, NY 11794-3840, USA}

\author{A.~Yu.~Smirnov}
\email{smirnov@ictp.trieste.it}
\affiliation{The Abdus Salam International Centre for Theoretical
  Physics, I-34100 Trieste, Italy}
\affiliation{Institute for Nuclear Research of Russian Academy of
  Sciences, Moscow 117312, Russia}

\begin{abstract}
    The measurement of the deviation of the 2-3 leptonic mixing from
    maximal, $D_{23} \equiv 1/2 - \sin^2\theta_{23}$, is one of the
    key issues for understanding the origin of the neutrino masses and
    mixing.  In the $3\nu$ context we study the dependence of various
    observables in the atmospheric neutrinos on $D_{23}$.  We perform
    the global $3 \nu$-analysis of the atmospheric and reactor
    neutrino data taking into account the effects of both the
    oscillations driven by the ``solar'' parameters ($\Dmq_{21}$ and
    $\theta_{12}$) and the 1-3 mixing. The departure from the
    one--dominant mass scale approximation results into the shift of
    the 2-3 mixing from maximal by $\Delta \sin^2\theta_{23} \approx
    0.04$, so that $D_{23} \sim 0.04 \pm 0.07 $ $(1\sigma)$. Though
    value of the shift is not statistically significant, the tendency
    is robust.  The shift is induced by the excess of the $e$-like
    events in the sub-GeV sample.  We show that future large scale
    water Cherenkov detectors can determine $D_{23}$ with accuracy of
    a few percent, comparable with the sensitivity of future long
    baseline experiments.  Moreover, the atmospheric neutrinos will
    provide unique information on the sign of the deviation (octant of
    $\theta_{23}$).
\end{abstract}


\maketitle


\section{Introduction}
\label{sec:intro}

The present $2\nu$ analysis of the atmospheric neutrino
results~\cite{SuperK,sk-nu2004,atmnp} in terms of
$\nu_\mu\leftrightarrow \nu_\tau$ oscillations gives as mass squared
difference and mixing:
\begin{equation} \label{eq:atmdata}
    \Dmq_{32} = (1.3 - 3.0) \times 10^{-3}~\eVq, \qquad
    \sin^2 2\theta_{23} \geq 0.94 \qquad \text{(90\% C.L.)} \,.
\end{equation}
The results of SOUDAN~\cite{SOUDAN} and MACRO~\cite{MACRO} experiments
are in a good agreement with~\eqref{eq:atmdata} and this oscillation
interpretation has been further confirmed by the K2K
results~\cite{K2K}.

The best fit of the data corresponds to maximal mixing $\sin^2
2\theta_{23} = 1.0$, and this is one of the most striking results in
neutrino physics.  The maximal or close to maximal mixing implies a
new symmetry of Nature which does not show up in other sectors of
theory indicating its non-trivial realization. 

However, at the moment, one cannot claim that the mixing is indeed
close to maximal. From the theoretical point of view the correct
parameter which characterizes the deviation is not $\sin^2 2
\theta_{23}$ but $\sin^2 \theta_{23}$ or 
\begin{equation} \label{eq:dev}
    D_{23} \equiv \frac{1}{2} - \sin^2 \theta_{23}.
\end{equation}
It is $\sin \theta_{23}$ that relates to the expansion parameter in
the neutrino mass matrix and $D_{23}$ characterizes violation of the
symmetry responsible for maximal mixing.  From~\eqref{eq:atmdata} we
obtain for the deviation 
\begin{equation} \label{eq:devb}
    |D_{23}| \leq 0.12, \qquad \text{(90\% C.L.)} \, , 
\end{equation}
and 
\begin{equation}
    (0.5 - \sin^2 \theta_{23}) / \sin^2\theta_{23} \sim 0.3. 
\end{equation}
That is, the deviation can be of the order of mixing itself. 

In the case of a significant deviation we cannot speak of a special
symmetry, and in fact, the large 2-3 mixing may appear as the sum of
small (order of Cabibbo) mixing angles (see, {\it e.g.},~\cite{DoSm}).
So, maximal or non-maximal mixing is equivalent to the dilemma of new
symmetry or no-new symmetry of Nature. (Here we exclude the 
possibility that small angles sum up to give accidentally exact
maximal mixing). Depending on value of the deviation the approach to
the underlying physics can be different.

The present data may already give some hint of deviation of the 2-3
mixing from maximal. Indeed, there is some excess of the $e-$like
events in the sub-GeV range. The excess increases with decrease of
energy within the sample~\cite{super-data-used}. In comparison with
predictions based on the atmospheric neutrino flux from
Honda~\cite{honda} the excess is about (12--15)\% in the low energy
part of the sub-GeV sample ($p < 0.4$ GeV, where $p$ is the momentum
of lepton) and it shows no significant zenith angle dependence. In the
higher energy part of the sub-GeV sample ($p > 0.4$ GeV) the excess is
about 5\%, and there is no excess in the multi-GeV region ($p > 1.33$
GeV).

In principle, the observed excess is within the estimated 20\%
uncertainty of the original atmospheric neutrino flux. So the $2\nu$
analysis of data with free overall normalization and tilt of the
energy spectrum can explain a large enough fraction of the excess
leading to the result of best fit maximal 2-3 mixing.

The excess has become more significant in the latest Super-Kamiokande
analysis~\cite{sk-nu2004}.  The recent data on primary cosmic 
rays~\cite{BESS,AMS} as well as the new 3-dimensional calculations of
the atmospheric neutrino fluxes~\cite{honda3d} imply a lower neutrino
flux, and therefore a larger excess which is becoming more difficult
to explain by a change of the atmospheric neutrino fluxes within their
expected uncertainties~\cite{super-data-used,sk-nu2004}.

Alternatively, such an excess can be explained (at least partly) by
the $\nu_e$-oscillations driven by the solar oscillation 
parameters~\cite{kim} provided that the 2-3 mixing deviates from
maximal~\cite{orl1,PeSm}. For the solar parameters which we will call
the LMA parameters the combined analysis of the solar~\cite{solar} and
KamLAND~\cite{kamland} data leads to 3$\sigma$
ranges~\cite{ourpostnu04}:
\begin{equation} \label{eq:solar}
    \Dmq_{21} = (7.4 - 9.2)\times 10^{-5}~\eVq, \qquad
    \tan^2 \theta_{12} =  0.28 - 0.58 \,.
\end{equation}
Oscillations of atmospheric neutrinos driven by these LMA parameters
have been widely discussed in the
literature~\cite{kim,orl1,PeSm,yasuda,sakai99,strumia,antonio,
Gonzalez-Garcia:2002mu}.  It is found that the relative change of the
atmospheric $\nu_e$ flux due to oscillations driven by the solar
parameters is determined by the two neutrino transition probability 
$P_2(\Dmq_{21}, \theta_{12})$ and a ``screening" factor~\cite{orl1}:
\begin{equation} \label{eq:fluxe}
    \frac{F_e}{F_e^0} - 1 =
    P_2 (\Dmq_{21}, \theta_{12}) \, (r \cos^2 \theta_{23} - 1) \,,
\end{equation}
where $F_e$ and $F_e^0$ are the electron neutrino fluxes with and
without oscillations, and $r \equiv F_{\mu}^0/F_e^0$ is the ratio of
the original muon and electron neutrino fluxes. The screening factor 
(in brackets) is related to the existence of both electron and muon 
neutrinos in the original atmospheric neutrino flux. 

In the sub-GeV region $r \approx 2$, so that the screening factor is
very small when the $\nu_{\mu} - \nu_{\tau}$ mixing is maximal. 
According to Eq.~\eqref{eq:fluxe}, the excess of the $e$-like events
can be written as:
\begin{equation} \label{eq:e-sub}
    \epsilon_e \equiv \frac{N_e}{N_e^0} -1  \approx
   \left(r D_{23} + \frac{r}{2}  -1 \right) \langle P_2 \rangle_{\nu 
\bar\nu},
\end{equation}
where $\langle P_2 \rangle_{\nu \bar\nu} \equiv [(1 - \xi) \langle P_2
\rangle + \xi \langle \bar{P}_2 \rangle ]$, and $\langle P_2 \rangle$
($\langle \bar{P}_2 \rangle$) is the average transition probability
for neutrinos (antineutrinos) in the Earth matter. The parameter $\xi$
gives the relative contribution of antineutrinos (without
oscillations). For the sub-GeV electrons we have $\xi\simeq 0.3$.
Once the solar oscillation parameters have been well determined, one
can calculate $P_2$ rather precisely. Then the study of the excess can
be used to search for the deviation $D_{23}$~\cite{PeSm}.

For the presently allowed range of solar oscillation parameters, 
neutrino oscillations can lead up to a (5--6)\% excess of the $e$-like
events in the sub-GeV atmospheric neutrino sample~\cite{orl1,PeSm}.
So, the oscillation explanation of the observed excess would imply
maximal allowed deviation $D_{23}$. On the other hand, the decrease of
$\sin^2 2\theta_{23}$ influences other observables (like high
statistics measurements of the zenith angle dependence of the
$\mu$-like events). Therefore to make a definitive conclusion about
the deviation one needs to perform a combined analysis of the all
available data and to take carefully into account the uncertainties in
the atmospheric neutrino fluxes.

The $\nu_e$ oscillations are also induced by non-zero 1-3 mixing and
$\Dmq_{31}$ responsible for the dominant mode of the atmospheric
neutrino oscillations. This effect is mostly visible for the multi-GeV
sample~\cite{ADLS,atmpetcov,palomares} for which the Earth matter
effect becomes important and can enhance the oscillations.  Non-zero
1-3 mixing induces also an interference effect in the sub-GeV
range~\cite{Gonzalez-Garcia:2002mu,PeSm}. However, within the present
bound on the 1-3 mixing from the CHOOZ reactor
experiment~\cite{CHOOZ}, the dominant factor which leads to a possible
excess of the sub-GeV e-like events is the $\Delta m^2_{21}$-driven
transitions discussed here.

In this paper we perform a detailed study of the dependence of the 
atmospheric neutrino observables on the deviation $D_{23}$. We
determine $D_{23}$ from the analysis of present data and investigate
possibilities of future experiments. 

The paper is organized as follows. In Sec.~\ref{sec:formalism} we
discuss the dependence of different samples of the atmospheric
neutrino data on $D_{23}$. In Sec.~\ref{sec:present} we present the
results of the global analysis of the atmospheric and CHOOZ results in
terms of three--neutrino oscillations where the effect of both mass
differences is explicitly considered.  In Sec.~\ref{sec:future} we
study the capabilities of future large scale water Cherenkov detectors
to determine $D_{23}$. Discussion of the results and conclusions are
given in Sec.~\ref{sec:conclusions}.


\section{$D_{23}$ and the atmospheric neutrino observables}
\label{sec:formalism}

In this section, using some approximate analytical results, we 
discuss the dependence of the atmospheric neutrino observables on the
deviation $D_{23}$.

\begin{enumerate}
  \item According to Eq.~\eqref{eq:e-sub} the excess of $e$-like events 
    in the sub-GeV range due to LMA parameter oscillations is
    proportional to the deviation $D_{23}$: 
    \begin{equation} \label{eq:e-suba}
	\epsilon_e \simeq D_{23}\, r\, \langle P_2 \rangle_{\nu \bar\nu},
    \end{equation}
    while their zenith angle distribution (encoded in $\langle P_2
    \rangle_{\nu \bar\nu}$) does not depend on $D_{23}$.  The 1-3
    mixing modifies this dependence but, for values compatible with
    the CHOOZ bound, it is a subdominant effect for sub-GeV events.
    
    The important point is that the excess decreases with energy as
    $\epsilon_e \sim E^{-2}$. This particular energy dependence of the
    excess allows to disentangle it from the uncertainties of the 
    neutrino fluxes. 
    
  \item As a consequence of this energy dependence, in the multi-GeV range 
    the excess of the $e$-like events due to the LMA parameters is
    very small: 5--10 times smaller than in the sub-GeV range and
    therefore below 1\%.  Conversely, the zenith angle distribution
    here is stronger. 
    
  \item The sub-GeV $\mu$-like events have more complicated dependence 
    on the deviation~\cite{orl1,Gonzalez-Garcia:2002mu}:
    \begin{equation} \label{eq:fluxmu1}
	\frac{N_{\mu}}{N_{\mu}^0} -1  =
	- \sin^2 2\theta_{23}
	\langle \sin^2 \frac{\phi}{2} \rangle_{\nu \bar\nu} - 
	\epsilon_{\mu} - \epsilon_{int} \,.
    \end{equation}
    The dominant contribution due to the $\nu_{\mu} - \nu_{\tau}$
    oscillations, $P_{\mu\tau}$, (the first term on the RHS) depends
    on $\sin^2 2\theta_{23} = 1 - 4 D_{23}^2$. Here $\phi$ is the
    oscillation phase due to $\Delta m^2_{31}$. The $\epsilon_{\mu}$
    term describes the decrease of the rate of the $\mu$-like events
    due to oscillations of the muon neutrinos into the electron
    neutrinos driven by the 1-2 mixing:
    \begin{equation} \label{eq:rel1}
	\epsilon_{\mu} \sim \frac{\cos^2\theta_{23}}{r}\epsilon_{e} 
	\approx D_{23} \cos^2\theta_{23}
	\langle P_2 \rangle_{\nu \bar\nu} \,.
    \end{equation}
    The third term is the interference of these two contributions and
    it is essentially averaged out. The expression \eqref{eq:fluxmu1}
    can be rewritten as
    \begin{equation} \label{eq:fluxmu2}
	\frac{N_{\mu}}{N_{\mu}^0} -1 \approx (4 D_{23}^2 -1) 
	\langle \sin^2 \frac{\phi}{2} \rangle_{\nu \bar\nu} 
	- D_{23} \cos^2\theta_{23} \langle P_2 \rangle_{\nu \bar\nu} \,.
    \end{equation}
    Notice that the 1-2 mixing has an opposite effect on the rate of
    the $\mu$-like events as compared to its effect on the e-like events.
    Thus for the case of the excess of $e$-like events, the 1-2 mixing
    diminishes the rate of $\mu$-like events.  So, it cancels
    partially the increase of the rate due to the main term 
    $P_{\mu\tau}$. Furthermore, both terms exhibit a different 
    dependence on $D_{23}$:
    \begin{equation} \label{eq:terms}
	P_{\mu\tau} \propto (1  - 4 D_{23}^2), \quad \epsilon_{\mu}  
	\propto D_{23}. 
    \end{equation}
    So, for large deviation the change of the main term dominates,
    whereas for small deviations the two contributions become
    comparable.
    
  \item For muons in the multi-GeV range, due to the suppression of
    $P_2$, the effect of the 1-2 mixing is small and the change of the
    main term in Eq.~\eqref{eq:fluxmu1} dominates down to very small
    deviations. So, basically the rate of the $\mu$-like events
    increases with $|D_{23}|$. 
    Conversely, for the multi-GeV events the effect of 1-3 mixing can
    be more substantial~\cite{ADLS,atmpetcov,palomares}.
    
  \item In the sub-GeV range the double ratio can be written as~\cite{orl1} 
    \begin{equation} 
	R_{\mu/e} = R_{\mu/e}^{\rm max}
	\frac{1 - 0.5 \sin^2 2\theta_{23} - {\epsilon}_{\mu}}
	{1 + {\epsilon_e}} \,,
    \end{equation}
    where $R_{e/\mu}^{\rm max}$ is the double ratio in the case of 
    two-neutrino oscillations with maximal mixing.  In terms of the
    deviation it can be rewritten as 
    \begin{equation} \label{eq:doublra}
        R_{\mu/e} = R_{\mu/e}^{\rm max}
        \frac{0.5 + 2 D_{23}^2  - D_{23} \cos^2 \theta_{23}
	  \langle P_2 \rangle_{\nu \bar\nu}}
	{1 + D_{23}\, r\, \langle P_2 \rangle_{\nu \bar\nu}} \,,
    \end{equation}
    For $D_{23} > 0$ the 1-2 mixing effects partially compensate the
    change of main term and numerical inspection of
    Eq.~\eqref{eq:doublra} reveals that the change of $R_{\mu/e}$ with
    $D_{23}$ is rather weak. For $D_{23} < 0$ the 1-2 mixing enhances
    the ratio $R_{\mu/e}$. 
    
    In the multi-GeV range the $\epsilon$ corrections are small and
    the double ratio increases with $D_{23}$. Comparing the double
    ratios in the sub-GeV and multi-GeV ranges we conclude that
    $R_{\mu/e} ({\rm sub-G})$ changes weakly (for $D_{23} > 0$)
    whereas $R_{\mu/e} ({\rm multi-G})$ increases, so that the ratio
    of ratios
    \begin{equation}
	R_{\mu/e} ({\rm multi-G})/R_{\mu/e} ({\rm sub-G})
    \end{equation}
    increases with $|D_{23}|$.
    
  \item For upward-going muons the average energy of the neutrinos is
    above 10 GeV.  For these energies, the effect of 1-2 mixing is
    strongly suppressed in matter. Also the possible 1-3 mixing has
    additional matter suppression: the relevant factor is $2EV/\Dmq
    \sim 0.1$.  So, to a good approximation the rate depends on $1 - 4
    D_{23}^2$.
\end{enumerate}

Summarizing, the rate of the low energy $e$-like events is
proportional to the deviation $D_{23}$ and the rates of high energy 
($\mu$-like) events depend on $(1 - 4 D_{23}^2)$, {\it i.e.}, on the
deviation squared. The rate of low energy $\mu$-like events an the
double ratios may have non-trivial interplay of the two dependences:
cancellation or enhancement of the main mode and 1-2 mixing
contributions.  With these consideration in mind one can better
understand the results of the various analyses described in the
following sections. 


\section{Global analysis of present data}
\label{sec:present}

\begin{figure} \centering
    \includegraphics[width=6in]{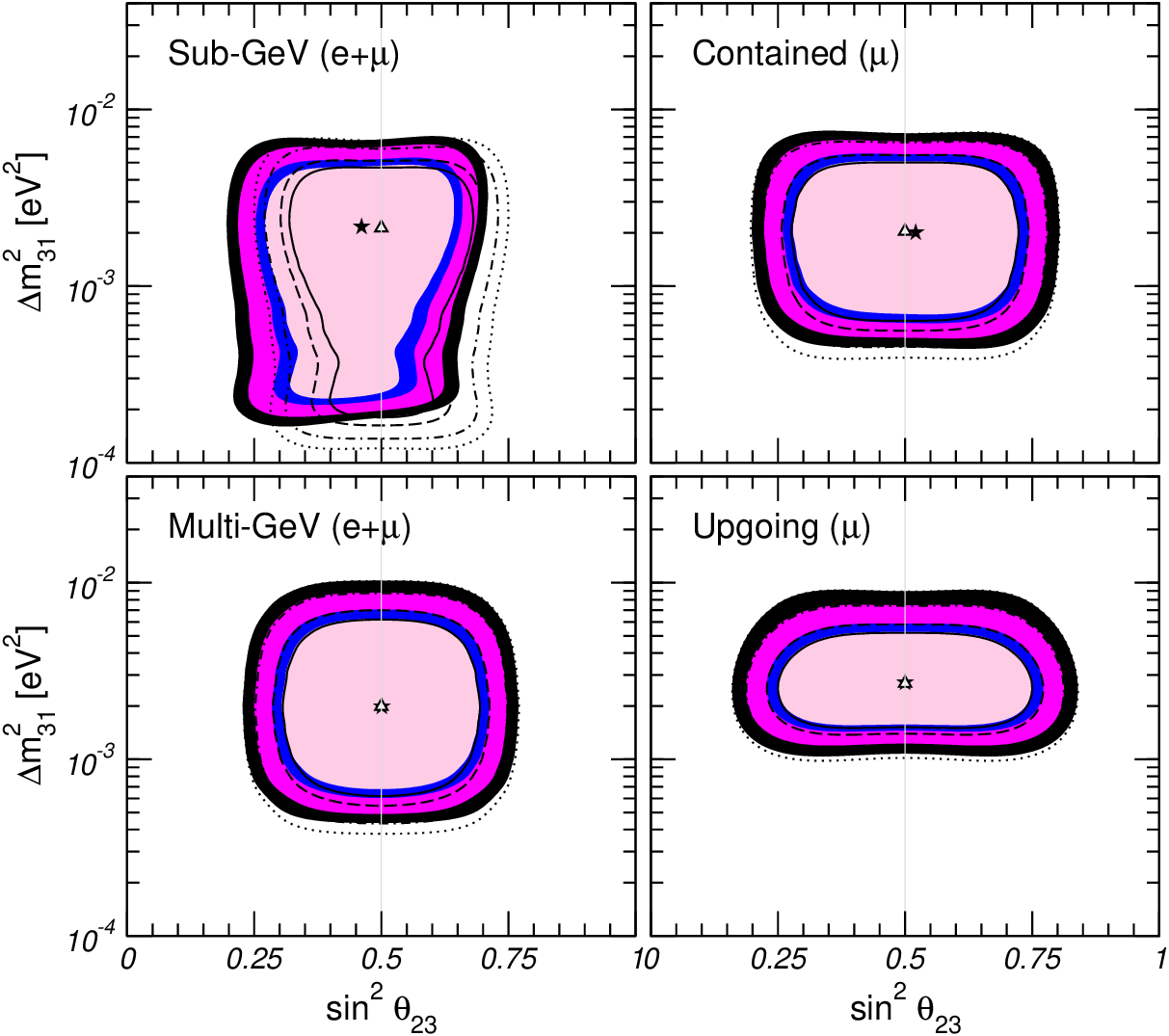}
    \caption{\label{fig:samples}%
      Allowed regions (90\%, 95\%, 99\% and $3\sigma$ C.L.) of the
      oscillation parameters $\Dmq_{13}$ and $\sin^2 \theta_{23}$ from
      the analysis of different atmospheric data samples. The best fit
      points are marked with either a star ($\Dmq_{21} \neq 0$) or a 
      triangle ($\Dmq_{21} = 0$) . Colored regions and stars
      correspond to $\Dmq_{21} = 8.2\times 10^{-5}~\eVq$ and
      $\tan^2\theta_{12} = 0.42$, whereas hollow regions and triangles
      are for $\Dmq_{21} = 0$. In both cases we assume $\theta_{13} =
      0$.}
\end{figure}

We discuss here what can be learned from the subleading effects
induced by non-vanishing solar splitting $\Dmq_{21}$ and the present
atmospheric and reactor neutrino data.  In our analysis we include the
complete 1489-day charged-current data set for SK-I~\cite{sk-nu2004},
which comprises the sub-GeV and multi-GeV $e$-like and $\mu$-like
contained event samples (each grouped into 10 bins in zenith angle),
as well as the stopping (5 angular bins) and through-going (10 angular
bins) up-going muon data events.  In the calculation of the event
rates we have used the new three-dimensional atmospheric neutrino
fluxes given in Ref.~\cite{honda3d}. 

Details of our statistical analysis based on the pull method can be
found in the Appendix of Ref.~\cite{atmnp} and here we summarize some
points which are essential for the present study. Together with the
statistical errors, we consider two types of uncertainties: the
theoretical and systematic ones. 

The theoretical uncertainties include uncertainties in the original
atmospheric neutrino fluxes and in the cross-sections. 

We have parametrized uncertainties of the atmospheric neutrino fluxes
in terms of four pulls: 

\begin{itemize}
  \item a total normalization error, which we set to 20\%; 
    
  \item a ``tilt'' factor which parametrizes possible deviations of
    the energy dependence of the atmospheric fluxes from the simple
    power law defined as 
    \begin{equation}
	\Phi_\delta(E) = \Phi_0(E) \left( \frac{E}{E_0} \right)^\delta
	\approx \Phi_0(E) \left[ 1 + \delta \ln \frac{E}{E_0} \right]
    \end{equation}
    with an uncertainty on the factor $\delta$, $\sigma_\delta=5\%$ 
    and $E_0 = 2$~GeV; 
    
  \item the uncertainty on the $\nu_\mu / \nu_e$ ratio, which is
    assumed to be $\sigma_{\mu/e}= 5\%$; and 
    
  \item the uncertainty on the zenith angle dependence which induces
    an error in the up/down asymmetry of events which we
    conservatively take to be 5\%. 
\end{itemize}

We also include independent normalization errors for the different
contributions to the interaction cross section: quasi-elastic
scattering (QE), $\sigma^{\sigma_\text{QE}}_\text{norm}=15\%$, single
pion production($1\pi$), $\sigma^{\sigma_{1\pi}}_\text{norm}=15\%$,
and deep inelastic (DIS) scattering (also refer to as multi-pion) for
which we estimate $\sigma^{\sigma_\text{DIS}}_\text{norm}=15\%$ for
contained events and $\sigma^{\sigma_\text{DIS}}_\text{norm}=10\%$ for
upward-going muons.\footnote{We also account for the uncertainty of
the $\sigma_{i,\nu_\mu} / \sigma_{i,\nu_e}$ ratio which is relevant
only for contained events, and it is much smaller than the total
normalization uncertainty.} 

We include as systematic uncertainties the experimental uncertainties
associated with the simulation of the hadronic interactions, the
particle identification procedure, the ring-counting procedure, the
fiducial volume determination, the energy calibration, the relative
normalization between partially-contained and fully-contained events,
the track reconstruction of upgoing muons, the detection efficiency of
upgoing muons, and the stopping-thrugoing separation.

In order to illustrate which data samples are more sensitive to the 
departure from the one--mass--scale dominance approximation and to the
deviation $D_{23}$ we first perform the analysis for different
sub-samples.  The results of these partial analysis are presented in
Fig.~\ref{fig:samples} where we show the allowed regions in the
($\Dmq_{31}$, $\sin^2\theta_{23}$) plane.  The colored (shadowed)
regions correspond to $\Dmq_{21} = 8.2\times 10^{-5}~\eVq$ and
$\tan^2\theta_{12} = 0.42$, whereas hollow regions are for $\Dmq_{21}
= 0$. In both cases we assume $\theta_{13} = 0$.  A comparison between
the two sets of regions clearly shows that the main effect of $\Delta
m^2_{21}$ oscillations appears in the $e$-like events at lower
energies as discussed in the previous section.  As seen in the figure
the inclusion of $\Dmq_{21}$-driven oscillations in the analysis
breaks the symmetry of $\theta_{23}$ around maximal mixing providing
the expected sensitivity to $D_{23}$.  In accordance with the
considerations of Sec.~\ref{sec:formalism} the allowed regions and the
best fit point shift to $\sin^2 \theta_{23} < 1/2$.  

Also, as discussed in the previous section, the effect is much more
suppressed at higher energies. As can be seen in the figure, the high
energy muon neutrino events as well as the multi-GeV ($e$-like and
$\mu$-like) events do not lead to the shift of 2-3 mixing from
maximal. The contained $\mu$-like events produce a very small shift in
the opposite direction: to $\sin^2 \theta_{23} > 1/2$. Let us
underline that despite they show much less sensitivity to 1-2
oscillations, the muon neutrino data are very important to constraint
the 2-3 oscillation parameters and to limit the effect of theoretical
and systematic uncertainties.

Next we study the effect of $\Dmq_{21}$ oscillation in the combined
analysis of all available atmospheric neutrino data (a total of 55
data points). In order to account also for the effect of the angle
$\theta_{13}$ we include in the analysis the results of the CHOOZ
experiment. For CHOOZ we consider the energy binned data.  This
corresponds to 14 data points (7-bin positron spectra from both
reactors, Table 4 in Ref.~\cite{CHOOZ}) with one constrained 
normalization parameter. In this analysis we have assumed CP
conservation but we have considered both possible values of the CP
parity which correspond to the CP phases $\delta=0$ and $\delta=\pi$.

\begin{figure} \centering
    \includegraphics[width=6in]{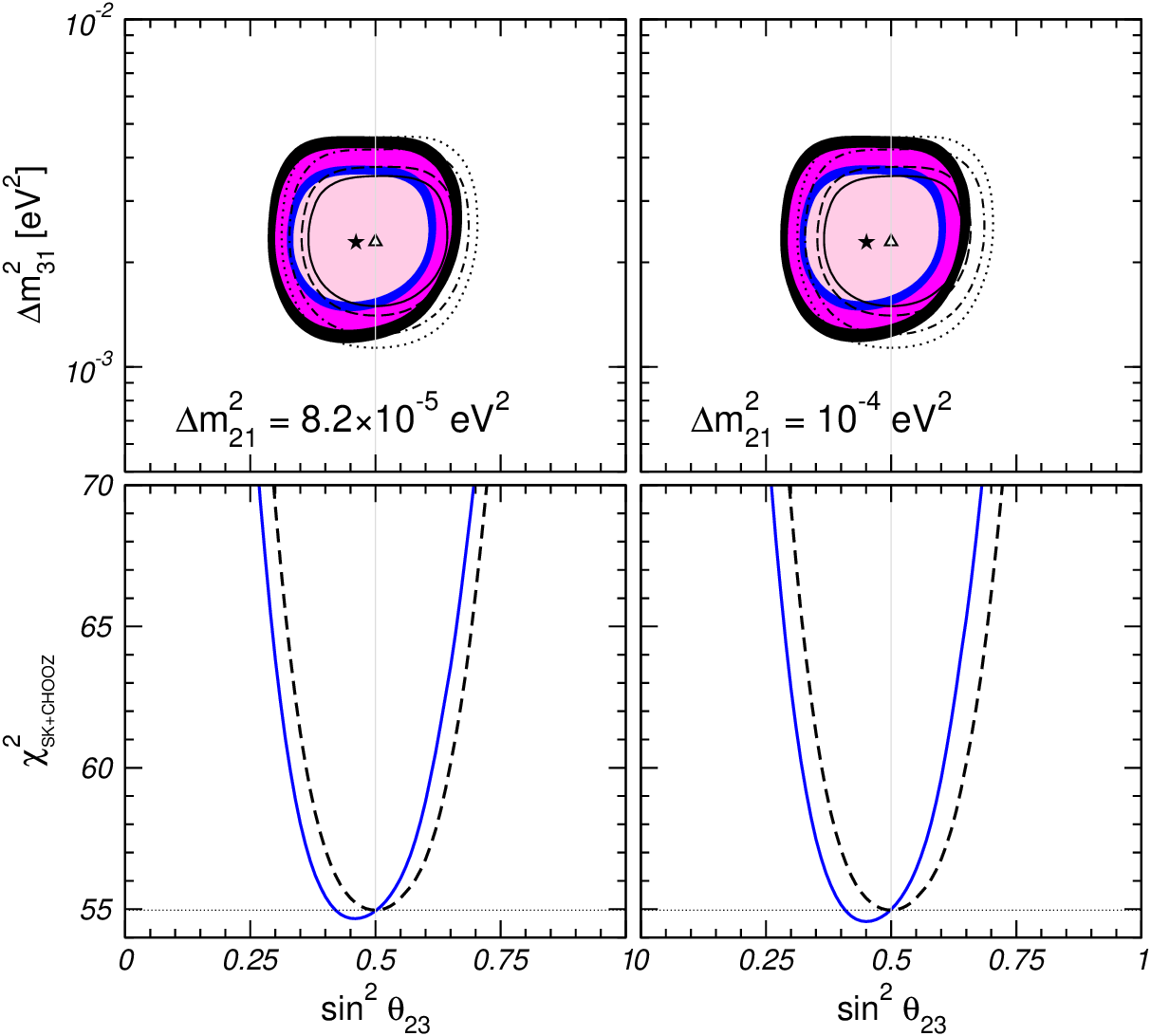}
    \caption{\label{fig:present}%
      Allowed regions (90\%, 95\%, 99\% and $3\sigma$ C.L.) of the
      oscillation parameters $\Dmq_{31}$ and $\sin^2\theta_{23}$ from
      the combined analysis of all the atmospheric and CHOOZ data
      samples. The best fit points are marked with either a star or a
      triangle. In the lower panels we show the dependence of the
      $\chi^2$ function on $\theta_{23}$, marginalized with respect to
      $\Dmq_{31}$. Colored regions, stars and solid blue lines
      correspond to $\tan^2\theta_{12} = 0.42$, $\theta_{13}$ free and
      $\Dmq_{21}$ set to the value indicated in each panel. Hollow
      regions, triangles and dashed black lines are for $\Dmq_{21} =
      0$ and $\theta_{13}$ free.}
\end{figure}

The results are shown in Fig.~\ref{fig:present}.  As before, the
colored (shadowed) regions correspond to $\tan^2\theta_{12} = 0.42$
and $\Dmq_{21}$ different from zero, whereas the hollow regions are
for $\Dmq_{21}= 0$. In order to verify explicitly that our results are
robust and do not change for non-zero $\theta_{13}$, we have
marginalized $\chi^2_{\rm SK+CHOOZ}$ with respect to this parameter.
In the lower panels we plot the $\chi^2$ function marginalized with
respect to $\Dmq_{31}$ as well.  

From the figure we see that, even with the present uncertainties, the
atmospheric data has some sensitivity to $\Dmq_{21}$ oscillation
effects and that these effects break the symmetry in $\theta_{23}$
around maximal mixing, although the effect is small.  Quantitatively,
for $\Dmq_{21} = 8.2\times 10^{-5}~\eVq$ and $\tan^2\theta_{12}=0.42$,
the best-fit point is located at $\sin^2\theta_{23} = 0.46$, with
$1\sigma$ ($3\sigma$) interval $(0.30)~0.39 \leq \sin^2\theta_{23}
\leq 0.54~(0.65)$ which means that for these values of the solar
parameters we find
\begin{equation} \label{eq:present}
    D_{23}=0.04\pm 0.07~ (^{+0.16}_{-0.19}) \,,
\end{equation}
whereas for $\Dmq_{21}=0$ we obtain $D_{23}=0.0\pm
0.07~(^{+0.18}_{-0.17})$ (the slight asymmetry of errors in this case 
is induced by $\theta_{13}$). For values of $\Dmq_{21}$ in the range
indicated by solar and KamLAND data, the deviation from maximal mixing
of $\theta_{23}$ increases with $\Dmq_{21}$. We have also verified
that, once CHOOZ is also included in the analysis, it makes little
difference to leave $\theta_{13}$ free or to set it to zero.

Let us stress that, although statistically not very significant, this
preference for non-maximal 2-3 mixing is a physical effect on the
present neutrino data, induced by the fact than an excess of events is
observed in sub-GeV electrons but not in sub-GeV muons nor, in the
same amount, in the multi-GeV electrons.  As a consequence, this
excess cannot be fully explained by a combination of a global
rescaling and a ``tilt'', of the fluxes within the assumed 
uncertainties. In the pull approach we find that both the total 
normalization and the tilt pulls are essentially fixed by the
combination of low-energy and high-energy muon data, and there is no
freedom left to accommodate the remaining excess of low energy
electron events.  Such an excess can only be partially explained by
means of another pull, the $\mu/e$ flavor ratio, whose uncertainty,
5\%, is however much smaller.  We have explicitly verified that in the
vicinity of the best-fit point the only pull which is affected by the
precise value of $D_{23}$ is the $\mu/e$ flavor ratio, whereas the
total normalization and the tilt are practically insensitive to it. 
In particular, we have checked that increasing the tilt uncertainty 
by a factor of 2 or allowing for a totally unconstrained overall
normalization does not affect the present value of $D_{23}$ in 
Eq.~\eqref{eq:present}.

From the lower panels in Fig.~\ref{fig:present} we can also see that
the quality of the fit slightly improves when $\Dmq_{21}$ differs from
zero. As expected, this is due to the fact that the non-vanishing 
value of $\Dmq_{21}$ and non-zero $D_{23}$ imply that $\nu_{\mu}
\rightarrow \nu_e$ transition is more efficient than the inverse one, 
$\nu_e \rightarrow \nu_{\mu}$, which allows to partially explain the
excess of $e$-like events observed by Super-Kamiokande in the sub-GeV
data sample. 

Notice that the central value of deviation in Eq.~\eqref{eq:present} 
corresponds to $\sin^2 2 \theta_{23} = 0.9936$ which is beyond the 
sensitivity of the next generation of the long-baseline experiments. 

In summary, in this section we have shown that atmospheric neutrino
data are sensitive to the subleading $\nu_\mu \to \nu_e$ conversion
induced by a non vanishing $\Dmq_{21}$. More important, the preference
of atmospheric data for maximal $\theta_{23}$ mixing appears to be a
specific property of the one--dominant mass scale approximation, and
seems to disappear when oscillations with the two wavelengths between
all three known neutrino flavor are considered. However, present data
still have far too little statistics to provide a conclusive answer.


\section{Sensitivity of future experiments}
\label{sec:future}

Having shown that atmospheric neutrino data can be a useful instrument
to search for deviations of $\theta_{23}$ from $45^\circ$, we now
discuss what can be learned from future atmospheric experiments. For
the sake of concreteness we have assumed a SK-like detector with
either 20 (\SKtim{20}) or 50 (\SKtim{50}) times the present SK-I
statistics and the same systematics as SK-I, and we have used the same
event samples as in SK. 

The procedure is as follows: First we simulate the signal according to
the expectations from some specific choice of the ``true'' values of 
parameters which we denote by $\Bomega$ 
\begin{equation}
    \Bomega \equiv ( \BDmq_{21} ,\, \BDmq_{31} ,\,
    \Btheta_{12} ,\, \Btheta_{13} ,\, \Btheta_{23} ) \,, 
\end{equation}
and then we construct
\begin{equation} \label{eq:chi2atm}
    \chi^2_\textsc{sk}(\Dmq_{21} ,\, \Dmq_{31} ,\,
    \theta_{12} ,\, \theta_{13} ,\,  \theta_{23} \,|\, \Bomega)
\end{equation}
assuming 20 or 50 times the present SK statistics and three choices
for the theoretical and systematic errors (see definitions in
Sec.~\ref{sec:present}):
\begin{itemize}
  \item[(A)] same theoretical and systematic errors as in present SK;
    
  \item[(B)] same systematic errors as in present SK, but no
    theoretical uncertainties;
    
  \item[(C)] neither theoretical nor systematic uncertainties (perfect
    experiment).
\end{itemize}
Next, in order to study the effect that non-zero values of
$\BDmq_{21}$ and $\Btheta_{12}$ can produce in the determination of
the atmospheric parameters $\Dmq_{31}$ and $\theta_{23}$ we define
\begin{multline} \label{eq:chq_atm}
    \chi^2_\Atm(\Dmq_{31} ,\, \theta_{23} \,|\, \Bomega)
    \equiv \min_{\Dmq_{21}, \theta_{13}}
    \Bigg[
    \chi^2_\textsc{sk}(\Dmq_{21} ,\, \Dmq_{31} ,\,
    \theta_{12} = \Btheta_{12} ,\, \theta_{13} ,\,  \theta_{23} \,|\, \Bomega)
    \\
    + \chi^2_\Chooz(\Dmq_{21} ,\, \Dmq_{31} ,\,
    \theta_{12} = \Btheta_{12} ,\, \theta_{13} \,|\, \Bomega) 
    + \Bigg( \frac{\Dmq_{21} - \BDmq_{21}}{\sigma_{\Dmq_{21}}} \Bigg)^2
    \Bigg]
\end{multline}
where we minimize with respect to the solar and reactor parameters
$\Dmq_{21}$ and $\theta_{13}$ and we keep only the explicit dependence
on the ``atmospheric'' parameters $\Dmq_{31}$ and $\theta_{23}$. The
assumption $\theta_{12} = \Btheta_{12}$ is made for purely practical
reasons because a complete scan of the whole five-dimensional
parameter space requires too much computer time. Note that regardless
of the specific assumptions on the `true values' $\BDmq_{21}$ and
$\Btheta_{13}$ the parameters $\Dmq_{21}$ and $\theta_{13}$ are
allowed to vary in our fit. 
In the definition of $\chi^2_\Atm$ in Eq.~\eqref{eq:chq_atm} we have
included also the CHOOZ experiment $\chi^2_\Chooz$ in order to have a
realistic bound on $\theta_{13}$. Similarly, the term $[( \Dmq_{21} -
\BDmq_{21}) / \sigma_{\Dmq_{21}})]^2$ accounts for the bound on
$\Dmq_{21}$ which is expected from KamLAND in the next few years.
Following Ref.~\cite{Bahcall:2003ce}, we have assumed that by then
$\BDmq_{21}$ will be known with an uncertainty of $3\%$ at $1\sigma$.
\begin{figure} \centering
    \includegraphics[width=5.6in]{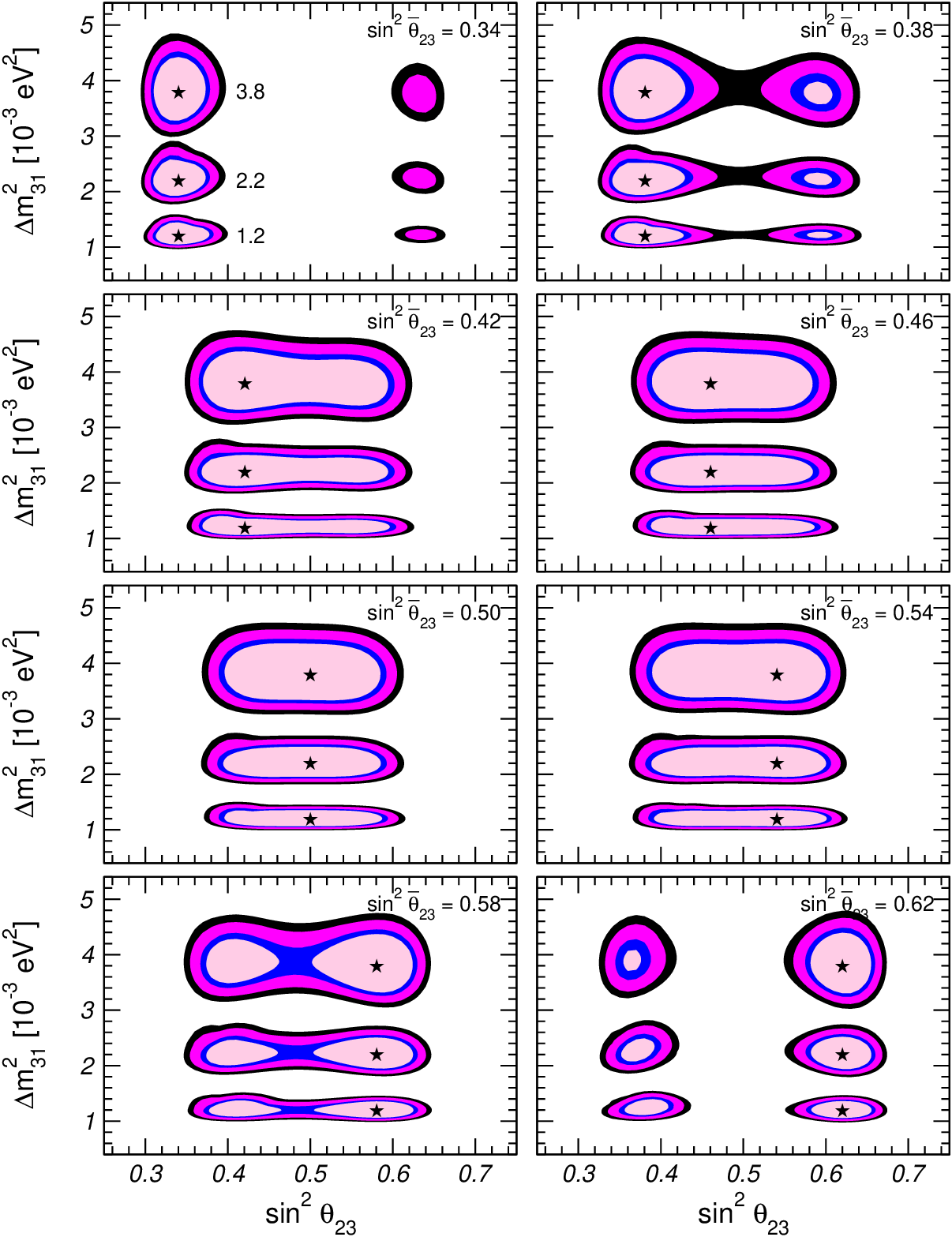}
    \caption{\label{fig:regions}%
      Allowed regions (at 90\%, 95\%, 99\% and $3\sigma$ C.L.) of
      oscillation parameters $\Delta m^2_{13}$ and $\sin^2\theta_{23}$
      expected from an atmospheric neutrino experiment with 20 times
      the present SK statistics and the same theoretical and
      systematic errors as in present SK. For definiteness, we choose
      $\Btheta_{13} = 0$, $\BDmq_{21} = 8.2\times10^{-5}~\eVq$ and
      $\tan^2 \Btheta_{12} = 0.42$, and we scan different values of
      $\BDmq_{31}$ and $\Btheta_{23}$. We also include the constraints
      from the CHOOZ experiment, as well as the sensitivity to
      $\Dmq_{21}$ expected after 3 years of KamLAND data
      (Eq.~\eqref{eq:chq_atm}). The undisplayed parameters $\Dmq_{21}$
      and $\theta_{13}$ are marginalized.}
\end{figure}

As an illustration of the expected sensitivity from future atmospheric
neutrino experiments, we show in Fig.~\ref{fig:regions} the allowed 
regions obtained from $\chi^2_\Atm$ assuming 20 times the present SK
statistics and the same theoretical and systematic errors as in
present SK (case A). For definiteness, we choose $\Btheta_{13} = 0$,
$\BDmq_{21} = 8.2\times10^{-5}~\eVq$ and $\tan^2 \Btheta_{12} = 0.42$,
and we scan different values of $\BDmq_{31}$ and $\Btheta_{23}$. From
this figure we find that:
\begin{itemize}
  \item future atmospheric neutrino experiments can observe and
    measure deviations of $\theta_{23}$ from maximal mixing, provided
    that $\theta_{23}$ is not too close to $45^\circ$: $\sin^2
    \theta_{23} < 0.38$ or $\sin^2 \theta_{23} > 0.60$; future
    reduction in the theoretical errors will further improve the 
    sensitivity; 
    
  \item they can discriminate between the ``light-side'' and
    ``dark-side'' for $\theta_{23}$, {\it i.e.}, they are sensitive to
    the octant of $\theta_{23}$.
\end{itemize}

In the rest of this section we quantify these two possibilities.


\subsection{Deviations from maximal mixing}

\begin{figure} \centering
    \includegraphics[width=6in]{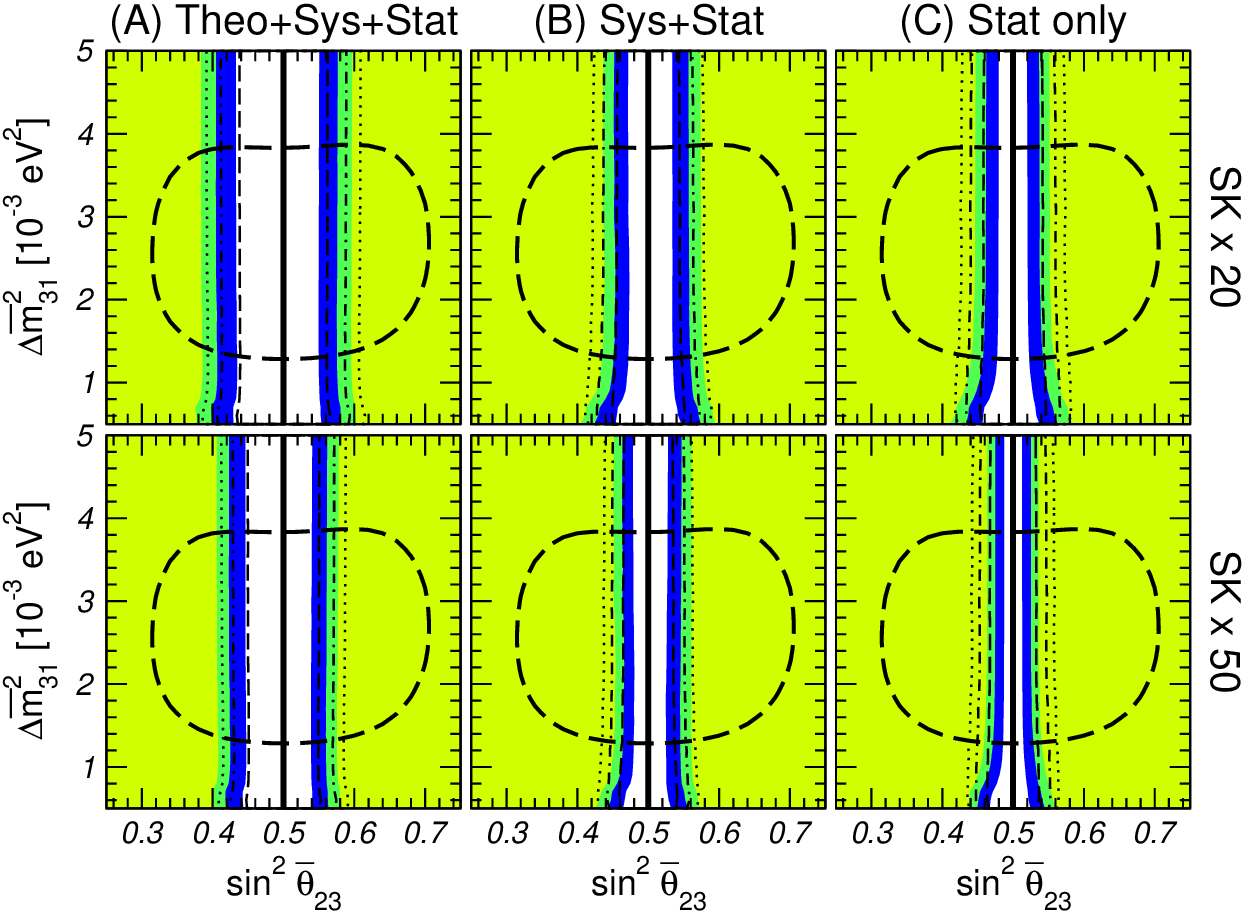}
    \caption{\label{fig:maximal}%
      ($\BDmq_{31}$, $\Btheta_{23}$) regions with 
      $\Delta\chi^2_\text{no-max}$ smaller than 1 (white), between 1
      and 4 (blue), between 4 and 9 (green), and larger than 9 (blue),
      respectively for $\BDmq_{21} = 8.2\times 10^{-5}~\eVq$ (the
      hollow regions are obtained for $\BDmq_{21} = 0$).  We set
      $\Btheta_{13} = 0$ and $\tan^2 \Btheta_{12} = 0.42$.  The dashed
      oval marks the $3\sigma$ region presently preferred by SK data.}
\end{figure}

Let us quantify the sensitivity of the future atmospheric neutrino
experiments to deviation of $\theta_{23}$ from $45^{\circ}$.  We
compare our results with the corresponding bounds which can be
expected for future long baseline (LBL)
experiments~\cite{Antusch:2004yx,minakata}, following the discussion
in Ref.~\cite{Antusch:2004yx}. The results of our analysis are
summarized in Fig.~\ref{fig:maximal} and Table~\ref{tab:maximal},
which can be directly compared with Fig.~1 and Table~1 of
Ref.~\cite{Antusch:2004yx}.

To perform this analysis, we have constructed the following function:
\begin{equation} \label{eq:chq_nomax}
    \Delta\chi^2_\text{no-max}(\Bomega)
    \equiv \min_{\Dmq_{31}, \theta_{23}} \bigg[
    \chi^2_\Atm(\Dmq_{31} ,\, \theta_{23} = 45^\circ \,|\, \Bomega) -
    \chi^2_\Atm(\Dmq_{31} ,\, \theta_{23} \,|\, \Bomega) \bigg]
\end{equation}
where $\chi^2_\Atm(\Dmq_{31} ,\, \theta_{23} \,|\, \Bomega)$ is given
in Eq.~\eqref{eq:chq_atm}. In Fig.~\ref{fig:maximal} we plot the
dependence of $\Delta\chi^2_\text{no-max}$ on $\BDmq_{31}$ and
$\Btheta_{23}$, for both $\BDmq_{21} = 8.2\times 10^{-5}~\eVq$
(colored regions) and $\BDmq_{21} = 0$ (hollow regions). We take
$\tan^2 \Btheta_{12} = 0.42$ and $\Btheta_{13} = 0$. The blue, green
and yellow regions correspond to $\Delta\chi^2_\text{no-max} > 1$, $4$
and $9$, respectively.  In other words, in Fig.~\ref{fig:maximal} we
display, for each value of $\BDmq_{31}$, the range of $\Btheta_{23}$
for which the simulated signal can be reconstructed as having maximal
$\theta_{23}$ at 1, 2 and 3 $\sigma$. The white region corresponds to
the the range of $\Btheta_{23}$ for which the simulated signal cannot
be distinguished from maximal $\theta_{23}$ at $1\sigma$. The
corresponding bounds on $\Btheta_{23}$ for $\BDmq_{31} = 2.2\times
10^{-3}~\eVq$ are summarized in Table~\ref{tab:maximal}.

\begin{table}
    \begin{tabular}{l@{\hspace{6mm}}c@{\hspace{3mm}}c@{\hspace{9mm}}c@{\hspace{3mm}}c}
	\hline
	\hline
	Experiment & \multicolumn{4}{c}{$|0.5 - \sin^2\Btheta_{23}|$}
	\\
	& \multicolumn{2}{c}{With $\BDmq_{21} = 8.2\times 10^{-5}~\eVq$}
	& \multicolumn{2}{c}{With $\BDmq_{21} = 0$} 
	\\
	& 90\% C.L. & $3\sigma$ & 90\% C.L. & $3\sigma$ \\
	\hline
	\SKtim{20} (A) Theo+Sys+Stat & [-0.086, 0.067] & [-0.116, 0.096] & [-0.080, 0.080] & [-0.108, 0.108] \\
	\SKtim{20} (B) Sys+Stat      & [-0.040, 0.050] & [-0.062, 0.075] & [-0.058, 0.058] & [-0.078, 0.078] \\
	\SKtim{20} (C) Stat only     & [-0.032, 0.032] & [-0.054, 0.052] & [-0.054, 0.054] & [-0.073, 0.073] \\
	\hline
	\SKtim{50} (A) Theo+Sys+Stat & [-0.070, 0.053] & [-0.094, 0.077] & [-0.064, 0.064] & [-0.087, 0.087] \\
	\SKtim{50} (B) Sys+Stat      & [-0.030, 0.040] & [-0.046, 0.061] & [-0.046, 0.046] & [-0.063, 0.063] \\
	\SKtim{50} (C) Stat only     & [-0.021, 0.021] & [-0.037, 0.036] & [-0.042, 0.042] & [-0.058, 0.058] \\
	\hline
	\hline
    \end{tabular}
    \caption{\label{tab:maximal}%
      Rejection of maximal mixing expected from future atmospheric
      neutrino experiments. We assume $\Btheta_{13} = 0$, $\tan^2
      \Btheta_{12} = 0.42$ and $\BDmq_{31} = 2.2\times 10^{-3}~\eVq$,
      and we study both the case $\BDmq_{21} = 8.2\times 10^{-5}~\eVq$
      and $\BDmq_{21} = 0$. The given intervals correspond to
      $\Delta\chi^2_\text{no-max}$ (see Eq.~\eqref{eq:chq_nomax})
      smaller than 2.71 (90\% C.L.) and 9 ($3\sigma$).}
\end{table}

From Fig.~\ref{fig:maximal} and Table \ref{tab:maximal} we find that
the sensitivity of atmospheric neutrino data to deviations from 
maximal mixing for large values of $\BDmq_{31}$ is comparable to what 
can be expected ``after ten years'' from LBL experiments according to
Ref.~\cite{Antusch:2004yx}, $D_{23}\leq 0.050\, (0.069)$ at 90\%
(3$\sigma$) CL.  Furthermore, for small values of $\BDmq_{31}$ the
atmospheric neutrino studies are much more sensitive than LBL
experiments, which lose sensitivity very fast when $\BDmq_{31}
\lesssim 2\times 10^{-3}~\eVq$ while the bound which can be obtained
from the atmospheric neutrino data is practically independent of the 
value of $\BDmq_{31}$.

The comparison among the left, central and right panels of
Fig.~\ref{fig:maximal} also shows that the sensitivity of atmospheric
neutrino data to deviations from maximal mixing improves considerably
if theoretical errors on the atmospheric fluxes and cross sections are
reduced. On the contrary setting to zero the systematic uncertainties
induce a smaller improvement. This implies that the obtained results
hold even if the future atmospheric neutrino experiment is affected by
somewhat larger systematics than the present SK detector has.

We also see that, as expected, when $\BDmq_{21}\neq 0$ the ranges of
$\Btheta_{23}$ can be asymmetric. This effect is mostly seen in the
first two panels (cases A and B) because larger errors allow for
larger values of $\overline{D}_{23}$.  We find that the overall effect
of the theoretical errors is such that the fit for maximal mixing is
``less-bad'' if an excess of e-like sub-GeV events is observed as
compared to the observation of a deficit, while for systematic
uncertainties the opposite holds.

In any case, comparing the solid (obtained with $\BDmq_{21} =
8.2\times 10^{-5}~\eVq$) and the hollow (obtained with $\BDmq_{21} =
0$) regions in Fig.~\ref{fig:maximal} we see that the value of the
solar mass splitting is not the most important effect in the
discrimination from maximal mixing, and the bound comes mainly from
muon data. Only when both theoretical and systematic uncertainties are
neglected (case C) the bound on $D_{23}$ becomes visibly sensitive to
$\Dmq_{21}$.  This occurs because the effect of a non-zero value of
$\Dmq_{21}$ is comparable to the small statistical error so this small
effect is relevant only when the fit is purely statistics-dominated.
In general, the subdominant $\Dmq_{21}$ effect is mostly important to
determine the octant of $\theta_{23}$ as we discuss next.


\subsection{Determination of the octant}

\begin{figure} \centering
    \includegraphics[width=6in]{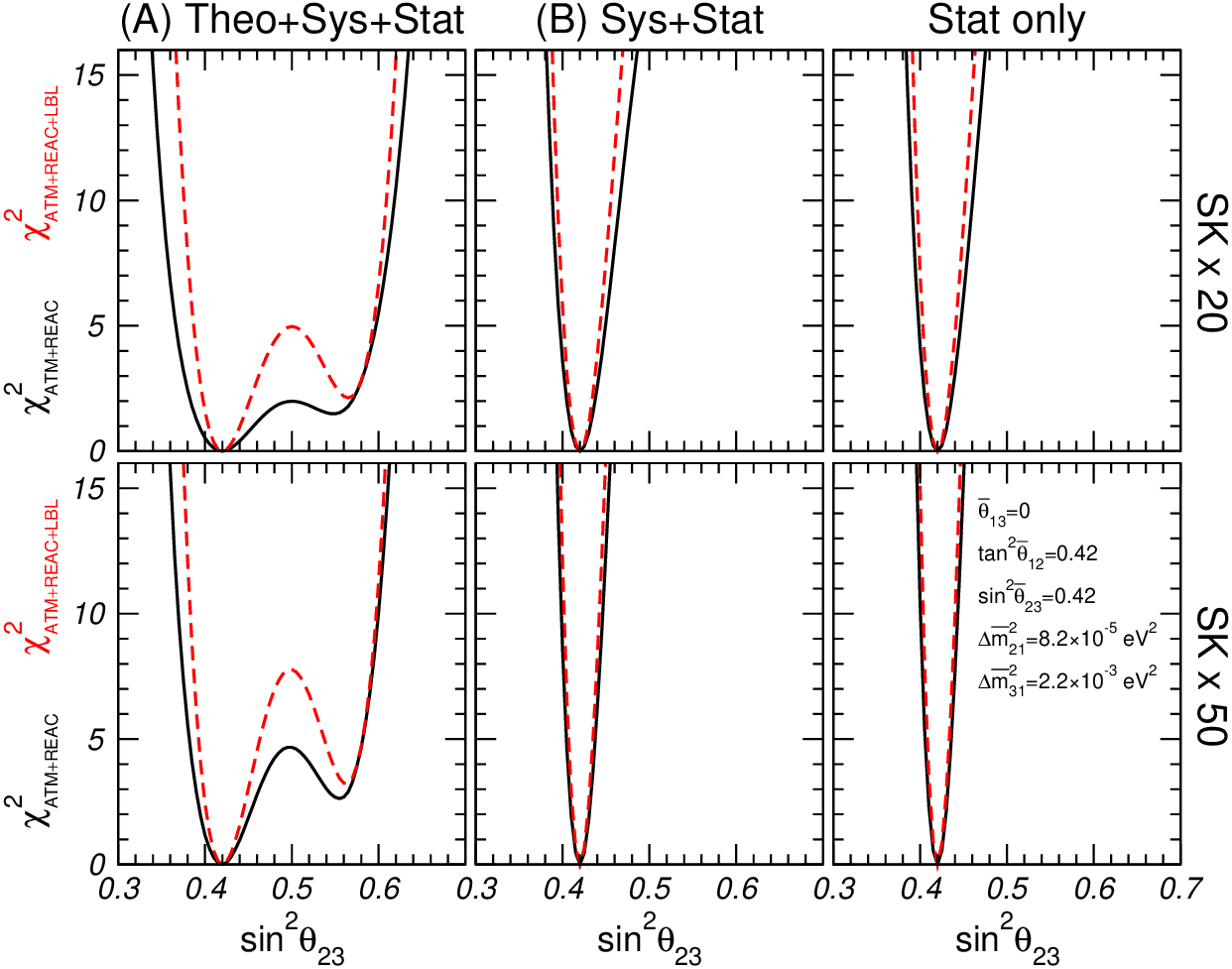}
    \caption{\label{fig:theta23}%
      Dependence of $\chi^2_\Atm$ (full black line) and $\chi^2_\Fut$
      (red dashed line) on $\sin^2\theta_{23}$, for $\Dmq_{31} =
      2.2\times 10^{-3}~\eVq$ and setting the simulated point
      ($\Bomega$) to $\Btheta_{13} = 0$, $\tan^2\Btheta_{12} = 0.42$,
      $\BDmq_{21} = 8.2 \times 10^{-5}~\eVq$, $\sin^2\Btheta_{23} =
      0.42$ and $\BDmq_{31} = 2.2\times 10^{-3}~\eVq$. For the
      definition of $\chi^2_\Atm$ and $\chi^2_\Fut$ see
      Eqs.~\eqref{eq:chq_atm} and \eqref{eq:chq_fut}.}
\end{figure}

As an illustration of the capability of atmospheric neutrino data to 
discriminate between $\theta_{23}$ smaller or larger than $45^\circ$,
we show in Fig.~\ref{fig:theta23} the dependence of $\chi^2_\Atm$ for
a particular simulated point $\Bomega$ as a function of $\theta_{23}$,
after marginalizing over all other parameters. In what follows we work
under the hypothesis that by the time this future atmospheric neutrino
experiment is in place we have found no-evidence of $\theta_{13}$ but
we may have a better-than-present determination of the oscillation
parameters from terrestrial experiments. To account for this effect we
have constructed the function
\begin{multline} \label{eq:chq_fut}
    \chi^2_{\Fut}(\Dmq_{31} ,\, \theta_{23} \,|\, \Bomega)
    \equiv \min_{\Dmq_{21}, \theta_{13}}
    \Bigg[
    \chi^2_\textsc{sk}(\Dmq_{21} ,\, \Dmq_{31} ,\,
    \theta_{12} = \Btheta_{12} ,\, \theta_{13} ,\,  \theta_{23} \,|\, \Bomega)
    \\
    + \chi^2_\Chooz(\Dmq_{21} ,\, \Dmq_{31} ,\,
    \theta_{12} = \Btheta_{12} ,\, \theta_{13} \,|\, \Bomega) 
    + \Bigg( \frac{\Dmq_{21} - \BDmq_{21}}{\sigma_{\Dmq_{21}}} \Bigg)^2
    \\
    + \Bigg( \frac{\Dmq_{31} - \BDmq_{31}}{\sigma_{\Dmq_{31}}} \Bigg)^2
    + \Bigg( \frac{\sin^22\theta_{23}-\sin^22\Btheta_{23}}{\sigma_{\sin^22\theta_{23}}} \Bigg)^2
    + \Bigg( \frac{\sin^22\theta_{13}-\sin^22\Btheta_{13}}{\sigma_{\sin^22\theta_{13}}} \Bigg)^2
    \Bigg]
\end{multline}
where in addition to the CHOOZ and KamLAND-3yr bounds we have included
a stronger bound on the $\theta_{13}$ angle (for example, from some
future reactor experiment~\cite{reactors}) as well as an improved
measurement of the atmospheric parameters $\Dmq_{31}$ and
$\theta_{23}$ from future narrow beam LBL experiments such as
T2K~\cite{t2k} or NuMi~\cite{numi}. Following
Ref.~\cite{Huber:2004ug}, we have assumed $\sigma_{\sin^22\theta_{13}}
= 0.01$, $\sigma_{\sin^22\theta_{23}} = 0.015$ and $\sigma_{\Dmq_{31}}
/ \BDmq_{31} = 0.015$. Note that in Eq.~\eqref{eq:chq_fut} we have 
expressed the sensitivity of future LBL experiments in terms of
$\sin^22\theta_{23}$, rather than $\sin^2\theta_{23}$, to account for
the fact that these experiments have no sensitivity to the
$\theta_{23}$ octant if $\theta_{13}$ turns out to be very small
(\emph{i.e.}, \ they cannot distinguish between $\sin^2\theta_{23} <
0.5$ and $\sin^2\theta_{23} > 0.5$)\footnote{A very long baseline wide
beam experiment such as the BNL proposal~\cite{bnl} could be also
sensitive to subdominant $\Delta m^2_{21}$
effects.}~\cite{Antusch:2004yx}.

In Fig.~\ref{fig:theta23} we see that the best fit point ($\chi^2=0$
by construction) is located at $\sin^2\theta_{23}^\text{true} =
\sin^2\Btheta_{23}$ (and $\Dmq_{31}= \BDmq_{31}$, $\Dmq_{21} =
\BDmq_{21}$, and $\theta_{13} = \Btheta_{13}$), while in the left
panels (case A) $\chi^2$ presents a second local minimum at
$\sin^2\theta_{23}^\text{false} \simeq 1 - \sin^2\Btheta_{23}$. The
shift between the position of the secondary minimum and the mirror
symmetric value of the true minimum is expected from the atmospheric
analysis.
On the other hand, in the central (case B) and right (case C) panels
the second ``false'' minimum has disappeared. Again, this illustrates
the importance of improving our knowledge of the fluxes and
cross-sections. As before, for both \SKtim{20} and \SKtim{50} the
dominant source of errors is the theoretical uncertainties, whereas
systematic uncertainties play a somewhat smaller role. We find that
once the theoretical uncertainties are neglected, the atmospheric data
can totally lift the degeneracy between the two octants of
$\theta_{23}$.

\begin{figure} \centering
    \includegraphics[width=6in]{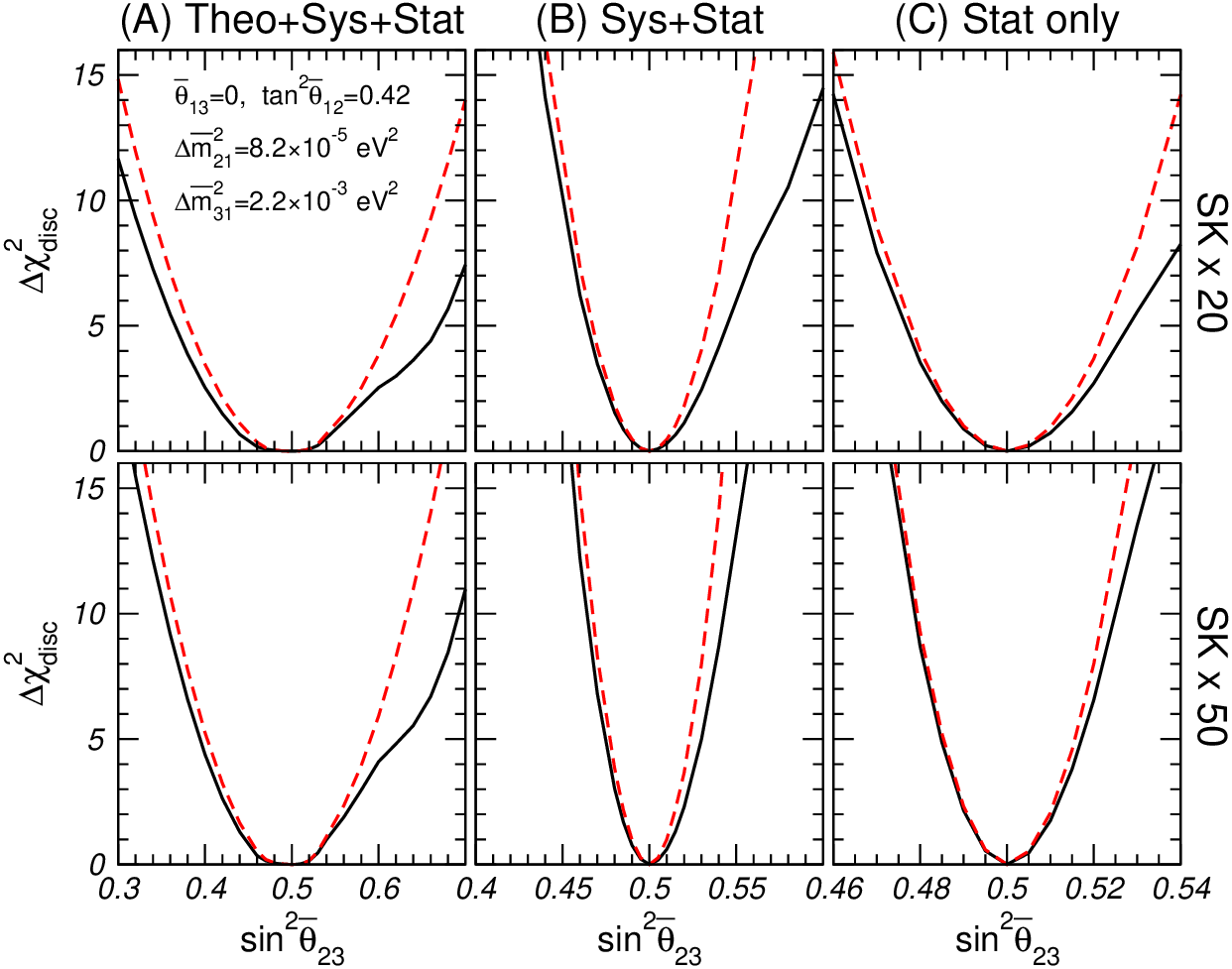}
    \caption{\label{fig:disc}%
      Dependence of $\Delta\chi^2_\text{disc}$ (see
      Eq.~\eqref{eq:dcq_disc}) on $\Btheta_{23}$, for $\Btheta_{13} =
      0$, $\tan^2\Btheta_{12} = 0.42$, $\BDmq_{21} = 8.2 \times
      10^{-5}~\eVq$ and $\BDmq_{31} = 2.2\times 10^{-3}~\eVq$. The
      different lines correspond to the same cases as in
      Fig.~\ref{fig:theta23}.}
\end{figure}

In order to quantify the discrimination power of the octant of
$\theta_{23}$ we define the difference 
\begin{equation} \label{eq:dcq_disc}
    \Delta\chi^2_\text{disc}(\Bomega)
    \equiv \min_{\Dmq_{31}} \Big[
    \chi^2_\textsc{atm+reac(+lbl)}(\Dmq_{31}, \theta_{23}^\text{false} \,|\, \Bomega) 
    \Big] - \min_{\Dmq_{31}} \Big[
    \chi^2_\textsc{atm+reac(+lbl)}(\Dmq_{31}, \theta_{23}^\text{true} \,|\, \Bomega)
    \Big]
\end{equation}
where $\theta_{23}^\text{false}$ is either the mixing angle of the
secondary local minimum, or $(90^\circ - \theta_{23}^\text{true})$ if
there is not a secondary local minimum. In Fig.~\ref{fig:disc} we plot
this difference as a function of the simulated ``true'' value 
$\sin^2\Btheta_{23}$ for $\tan^2\Btheta_{12} = 0.42$, $\BDmq_{21} =
8.2 \times 10^{-5}~\eVq$ and $\BDmq_{31} = 2.2\times 10^{-3}~\eVq$.
This function is very insensitive to the exact value of $\BDmq_{31}$
in the interval $10^{-3}~\eVq \leq \BDmq_{31} \leq 5 \times
10^{-3}~\eVq$.

The figure shows that unless $\theta_{23}$ is very close to maximal
mixing such a future atmospheric neutrino experiment can provide a
meaningful determination of the octant of $\theta_{23}$.  For example,
from the figure we read that for SK$\times$50 the octant of
$\theta_{23}$ can be determined at 90\% CL if
\begin{align}
    \sin^2\theta_{23} & \leq 0.42~[\theta_{23}\leq 40^\circ] & & \text{or} &
    \sin^2\theta_{23} & \geq 0.57~[\theta_{23}\geq 49^\circ] & & \text{(A)} \nonumber
    \\
    \sin^2\theta_{23} & \leq 0.48~[\theta_{23}\leq 43^\circ] & & \text{or} &
    \sin^2\theta_{23} & \geq 0.52~[\theta_{23}\leq 46^\circ] & & \text{(B)}
    \\
    \sin^2\theta_{23} & \leq 0.49~[\theta_{23}\leq 44.4^\circ] & & \text{or} &
    \sin^2\theta_{23} & \geq 0.51~[\theta_{23}\leq 45.6^\circ] & & \text{(C)} \nonumber
\end{align}
These results are almost independent of the exact value of
$\BDmq_{31}$ within the present atmospheric region. From
Fig.~\ref{fig:disc} we see that the discriminating power can be
slightly improved if LBL experiments provide a better determination of
$|D_{23}|$, as a consequence of the shift in the position of the
secondary minimum in the atmospheric neutrino analysis.

A final comment on the role of $\theta_{13}$. If $\Btheta_{13}$ is not
very small, the oscillation probabilities at the considered LBL
experiments are also not symmetric under the change of octant and they
can also contribute to the octant discrimination~\cite{minakata}.  
This effect has not been statistically quantified in detail in the
literature and it is beyond the purpose of this paper.


\section{Conclusion}
\label{sec:conclusions}

In this paper we have discussed the phenomenology of atmospheric 
neutrinos associated with the deviation of the 2-3 leptonic mixing
from maximal.  Our main results can be summarized as follows:
\begin{enumerate}
  \item We have performed the global $3\nu$-analysis of the
    atmospheric and reactor neutrino data taking into account the
    effect of both the oscillations driven by the ``solar'' parameters
    ($\Dmq_{21}$ and $\theta_{12}$) and of the 1-3 mixing. The results
    are shown in Fig.~\ref{fig:present}. We find that the departure
    from the one--dominant mass scale approximation in the analysis
    results into the shift of the 2-3 mixing from maximal, so that
    $D_{23} \sim 0.04 \pm 0.07 $. The shift is due to the excess of
    $e$-like events in the sub-GeV sample as illustrated in 
    Fig.~\ref{fig:samples}. For these values of $D_{23}$ the LMA
    oscillations explain the excess only partly.  Larger deviation is
    disfavored by the zenith angle distribution of the $\mu$-like
    events. The qualitative effect of the shift of 2-3 mixing from
    maximal one is robust.  Though particular value of the shift
    depends on details of the treatment of errors.  
    
  \item Future experiments will have much higher sensitivity to
    $D_{23}$.  With $20 - 50$ SK statistics and better knowledge of
    the cross-sections and the original fluxes the atmospheric
    neutrinos will probe $D_{23}$ down to few percent (see
    Fig.~\ref{fig:maximal} and Table~\ref{tab:maximal}) -- a
    sensitivity comparable with that attainable at future LBL
    experiments. This sensitivity does not change with decrease of
    $\Dmq_{31}$ and therefore a high statistics atmospheric neutrino
    experiment is better than LBL experiments to determine deviations
    of $\theta_{23}$ from maximal mixing if $\Dmq_{31}$ lies in the
    lower part of the present allowed range.
    
  \item If $D_{23} \neq 0$ future atmospheric neutrino experiments
    have the potentiality to discriminate the octant due to effects 
    associated to the LMA oscillations, as shown in 
    Figs.~\ref{fig:theta23} and~\ref{fig:disc}.
\end{enumerate}


\acknowledgments

M.M.\ is grateful to P.~Huber and T.~Schwetz for very useful
discussions on long-baseline experiments.  A.Yu.S.\ thanks P.\ Lipari
for valuable comments.  
This work was supported in part by the National Science Foundation
grant PHY-0354776. M.C.G.-G.\ is also supported by Spanish Grants No
FPA-2001-3031 and CTIDIB/2002/24.


\end{document}